# Efficient Upload Bandwidth Estimation and Communication Resource Allocation Techniques

MUGUREL IONUT ANDREICA, NICOLAE TAPUS
Computer Science and Engineering Department
Politehnica University of Bucharest
Splaiul Independentei 313, sector 6, Bucharest
ROMANIA
{mugurel.andreica, nicolae.tapus}@cs.pub.ro   https://mail.cs.pub.ro/~mugurel.andreica

*Abstract:* - In this paper we address two problems, for which we present novel, efficient, algorithmic solutions. The first problem is motivated by practical situations and is concerned with the efficient estimation of the upload bandwidth of a machine, particularly in the context of a peer-to-peer content sharing and distribution application. The second problem is more of a theoretical nature and considers a constrained communication resource allocation situation.

*Key-Words:* - Upload bandwidth estimation, Communication, Resource allocation, Peer-to-peer, Cooperative method.

## 1 Introduction

Communication is a key topic in every distributed system. Providing communication Quality-of-Service (QoS) guarantees or becoming aware of current communication parameters are two important tasks nowadays. In this paper we tackle two problems: a very practical problem, concerned with the estimation of the upload bandwidth of a machine, and a theoretical problem, concerned with the analysis of constrained communication resource allocations. The rest of this paper is structured as follows. In Section 2 we discuss the problem of estimating the upload bandwidth of a machine and we present our proposed solution. In Section 3 we discuss a theoretical communication resource allocation problem, for which we provide novel algorithmic solutions and we identify new patterns. In Section 4 we present related work and we conclude.

## 2 Upload Bandwidth Estimation

Estimating the upload bandwidth of a machine (e.g. computer) is extremely useful in a wide variety of scenarios and applications, like, for instance, peer-to-peer applications based on the Bittorrent tit-for-tat mechanism or other similar techniques (many file sharing, live streaming, and video on demand systems belonging to this class have been proposed during the past few years [2, 3]). In such systems, the downloaded data of every peer *P* is proportional to the data uploaded by peer *P* to the other peers. Since in order to maximize its overall utility, a peer wants to download data at high transfer rates, it must also be able to upload data to other peers at high speeds. However, most Internet users are connected to the Internet via asymmetric links, in which the download speed (bandwidth) is significantly higher than the upload speed (bandwidth). As such, the situation in which the upload bandwidth is fully utilized can easily occur. Such a situation may cause some problems. One of the most pregnant ones is the behavior of TCP flows when the upload link is congested. Through experiments, we determined that if a peer *P* downloads data at a rate *D* through a TCP connection, then an upload rate *U* of up to *2-5%* of *D* is used by the TCP protocol for sending *ACK* messages. If the upload link is congested and less than *U* bandwidth is available, the download rate *D* cannot be maintained and the TCP protocol makes use of its well known AIMD mechanism, which reduces the download speed drastically in a short time (while allowing it to increase back to its former values only slowly). Thus, when the upload link is congested, the download rates of TCP connections are, on average, far from the optimal performance. If, however, we knew the (available) upload bandwidth, we could reserve part of it for TCP acknowledgements, thus maintaining the download rate at a high average value. Other situations in which knowing the (available) upload bandwidth of a machine is useful are concerned with the implementation of higher-level functions and behaviors, like content seeding, peer selection, bandwidth trading, and so on. In this section we will present a novel upload bandwidth estimation technique, which was partly developed in the context of the European Union FP7 project P2P-Next. At the moment, the technique is applicable for estimating the upload capacity of a machine (i.e. its total upload bandwidth), in the absence of background traffic. The technique also works when background traffic is present, but it does not compute the available upload bandwidth, because the background communication flows can be influenced by our method.

An upload bandwidth estimation technique should be as non-intrusive as possible (i.e. it should generate little extra traffic). If possible, it would be desirable to make use of the existing traffic in order to estimate the upload bandwidth. Due to portability reasons, the technique should be implemented in user-space and should not make use of operating system-specific functions.

Our proposed technique works as follows. When a peer $S$ wants to estimate its upload bandwidth, it will need the help of $N \geq 1$ other helper peers ($P(1), ..., P(N)$). Peer $S$ will send $M(i)$ packets to each peer $P(i)$ ($1 \leq i \leq N$). The packets sent to the same peer $P(i)$ must have equal sizes ($PSize(i)$), but packets sent to different peers may have different sizes. It is also not necessary to send the same number of packets to every peer $P(i)$. Peer $S$ will send the $M(1)+...+M(N)$ packets one after another, in some order, such that any 2 consecutive packets sent by $S$ should preferably be sent to two different peers. What is important, however, is that the upload bandwidth of the peer $S$ should be constantly used, i.e. there should be no delays between two consecutive packets sent by $S$. We assume a FIFO queue at the sender (as is usually the case), i.e. the packets are transferred on the upload link in the order in which they are sent by $S$ (no matter to which helper peer they are sent).

When a peer $P(i)$ receives the $j^{th}$ packet, this packet will also contain the value $TAB(i,j)$=the total amount of bytes that peer $S$ has sent to all the $N$ peers up to the moment when the currently received packet was sent by $S$ (including the size $PSize(i)$ of the currently received packet). Then, let $TAB(i,j-1)$ be the value received by $P(i)$ at the previous packet (we consider the case $j \geq 2$). Let's assume that packet $j-1$ was received by $P(i)$ at time $T(i,j-1)$ and packet $j$ was received at time $T(i,j)$. Peer $P(i)$ will compute an estimation $U(i,j-1)=(TAB(i,j)-TAB(i,j-1))/(T(i,j)-T(i,j-1))$ of the upload bandwidth of peer $S$. Note that some of the packets sent by peer $S$ may be lost and the $j^{th}$ packet received by peer $P(i)$ may not necessarily be the $j^{th}$ packet sent by peer $S$ to $P(i)$. The packets may also be received out of order. When a peer $P(i)$ receives a packet, it first checks if the information contained in the packet regarding the total number of bytes sent so far by peer $S$ is larger than that of the previously received packet (unless it is the first received packet) - if the information value is not larger, then the currently received packet is discarded. The information regarding the total number of bytes sent by peer $S$ acts as a sequence number for the packets, because it increases with time. After sending the last packet to every peer $P(i)$, peer $S$ notifies every peer $P(i)$ that the test is complete (the notification should preferably not be lost, although it is not important if a small fraction of peers do not receive the notification). Every peer $P(i)$ has a time limit for waiting for new packets. When this limit is exceeded, it will assume that the test is complete (i.e. it will behave as if it had received the test completion notification). At the end, every peer $P(i)$ has $E(i)$ estimations: $U(i,1), ..., U(i,E(i))$. We will remove from this set the outliers (the values which are too high or too low) and compute an average $Uavg(i)$ of the remaining values. Peer $P(i)$ will then send $Uavg(i)$ to peer $S$. For the outliers removal we considered the following technique. We compute the median value $U_{med}$ of the estimations. Then, we remove all the estimations which are smaller than $p_1 \cdot U_{med}$ or larger than $p_2 \cdot U_{med}$ (for some carefully chosen values $0 \leq p_1 \leq 1$ and $p_2 \geq 1$). Afterwards, we perform an iterated removal of borderline values. As long as we have more than $K$ estimations left (e.g. $K=3$) we perform the following action: (1) we compute $U_m$=the average of the values of the remaining estimations and $sgm$=the standard deviation; (2) we remove all the estimations whose values do not belong to the interval $[U_m - q \cdot sgm, U_m + q \cdot sgm]$ (for a carefully chosen value of $q$; e.g. $q=1$); (3) if no values were removed in step (2) then we break the loop. In the end, peer $S$ will receive the estimations $Uavg(i)$ from (some of) the peers $P(i)$. If at least a fraction $PA$ (e.g. $PA=0.6$) of these values are "close" (and at least $PB \cdot N$ values were received; $0 < PB \leq 1$), then we remove the other values and compute the average of the remaining values: this will be the estimated upload bandwidth. We define *closeness* as follows. We compute the median $U_{md}$ of the received values and then we compute the number of received values which lie in the interval $[p_3 \cdot U_{md}, p_4 \cdot U_{md}]$ (where $0 \leq p_3 \leq 1$ and $p_4 \geq 1$). If we do not have at least a fraction $PA$ of "close" values, then it is possible for the estimated values to be too low, because the upload bandwidth estimations of peer $P(i)$ are also influenced by the available bandwidth $AB(S,P(i))$ of the path between $S$ and $P(i)$ (in fact, theoretically, we have $T(i,j)-T(i,j-1)=max\{(TAB(i,j)-TAB(i,j-1))/SUB, PSize(i) / AB(S, P(i))\}$, where $SUB$ is the upload bandwidth of peer $S$). This issue can be solved by sending larger packets or by using more helper peers: this way, two consecutive packets will reach a peer $P(i)$ after a larger time interval, overcoming the influence of $AB(S,P(i))$. Fig. 1 depicts the proposed technique, in which the same number of equally sized packets is sent to each of the $N=4$ helper peers in a round-robin fashion.

Let's have a closer look now at the way the upload bandwidth estimation technique works. If $N=1$, then $P(1)$ actually estimates a value $B$ which is upper bounded by the smaller of the following two values: the (available) bandwidth of the path between $S$ and $P(1)$ and the upload bandwidth of $S$. In fact, it is possible that the available bandwidth from peer $S$ to any of the helper peers is smaller than the upload bandwidth of peer $S$. However, by sending packets to multiple peers (e.g. in a round-robin fashion), peer $S$ does not congest the paths to the helper peers. Moreover, a helper peer $P(i)$ also receives the total number of bytes sent by peer $S$ so far during the test. The difference between the total number of bytes transmitted with two consecutive packets received by $P(i)$ is larger than the number of bytes that peer $S$ could have sent directly to $P(i)$ in the same time period (when $N>1$). A requirement for the technique to work correctly is that the sum of the bandwidths of the paths from $S$ to every helper peer should be at least as large as the upload bandwidth of peer $S$ (that is why larger packets or more helper peers are useful). If, however, the paths from $S$ to multiple helper peers share common bottleneck links (other than the upload link of peer $S$), then the technique may still incorrectly estimate (e.g. underestimate) the upload bandwidth. It is, thus, desirable for the helper peers to be geographically distributed, so that the paths from $S$ to the helper peers may be as disjoint as possible. Let's notice that we can use the method presented above for estimating the upload bandwidth in a continuous manner. Peer $S$ repeatedly sends packets to each of the peers $P(i)$. After receiving the $j^{th}$ packet ($j \geq MinP(i)$), peer $P(i)$ can provide an estimation $Uavg(i,j)$ using the previous $MinP(i)$ estimations (thus, we use a sliding window kind of approach). $MinP(i)$ is the minimum number of packets peer $P(i)$ needs to receive in order to consider the estimations to be statistically relevant.

If the upload bandwidth estimation technique is used in order to help the decision making process of an application $App$, then the technique can make use of the information regarding the upload bandwidth consumed by $App$ (this information can be made available, as the technique is integrated into $App$). We will consider that all the upload traffic generated by $App$ is sent to a "virtual helper peer" $P(N+1)$, which will not send any estimation back (although, in reality, the upload traffic of $App$ may have multiple destinations). Note that in order for the presented technique to produce reliable results, peer $S$ must never be idle in terms of upload traffic (i.e. it should always upload something): this is because when a peer $P(i)$ measures the time difference between two consecutive packets that it receives, it makes the implicit assumption that peer $S$ has been uploading data during all this time. Thus, as long as the upload buffer(s) of peer $S$ are not empty, peer $S$ does not have to send a new packet to any of its helper peers; it just has to increase the counter of the total number of bytes sent by $S$. However, since upload bandwidth estimations are received by peer $S$ only from the peers $P(i)$, peer $S$ cannot postpone indefinitely the sending of a packet to a peer $P(i)$ (even if the application $App$ generates enough traffic). Thus, the previously described technique can be modified as follows. Peer $S$ will only send the next packet to the next peer $P(i)$ (e.g. in the round-robin order) if the amount of upload buffer space used by $App$ is below a certain threshold or the duration between the moment when the previous message was sent to a helper peer $P(*)$ and the current time moment exceeds a given time limit. Note that when we also use existing $App$ traffic, we can reduce the number of packets after which a peer $P(i)$ computes an estimation, thus reducing the overall extra traffic generated by this method. The number of packets after which an estimation is computed could even be determined by each helper peer separately, based on the values $(TAB(i,j)-TAB(i,j-1))$ and $(T(i,j)-T(i,j-1))$ (e.g. the larger these values are, the fewer packets are required before obtaining an accurate estimation).

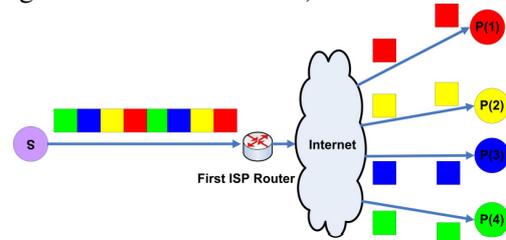

Fig. 1. Upload bandwidth estimation in progress.

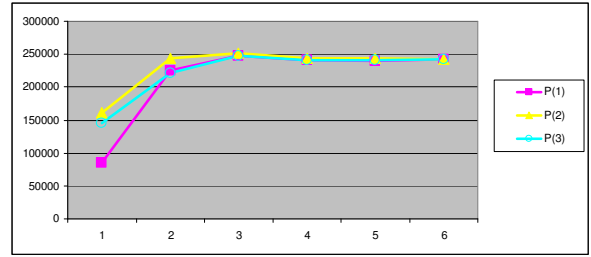

Fig. 2. Estimated upload bandwidth (Bps) as a function of packet size ($2^{10}$-$2^{15}$ bytes).

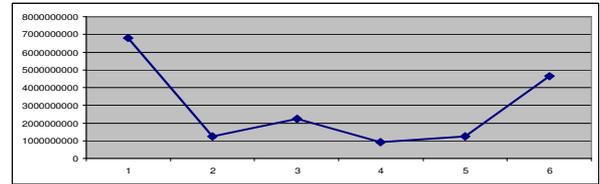

Fig. 3. The product between the error of the estimation and the total generated traffic for each of the *6* tests.

We implemented the proposed technique in the Python programming language and we validated it as follows. The source peer $S$ was located behind a NAT (in Bucharest), running Windows Vista; we used $N=3$ helper peers, all running Linux: $P(1)$ was located at the Technical University of Delft (Netherlands), $P(2)$ was located at the University of Craiova (Romania), and $P(3)$ was located at the Politehnica University of Bucharest (Romania). We sent $M=20$ packets to every helper peer, in a round-robin manner; we used $p_1=0.2$, $p_2=5$, $p_3=0.8$, $p_4=1.2$, $PA=PB=0.6$ and $PSize(1)=…=PSize(N)$. We ran *6* tests, in which we only changed the packet sizes: *1024*, *2048*, *4096*, *8192*, *16384*, and *32768* bytes. The source peer opened one TCP connection to every helper peer, for sending the corresponding packets (we designed our own protocol in order to mark the beginning and the end of a packet). The estimation computed by each helper peer is presented in Fig. 2. Later, we also performed the same test, by sending UDP packets (instead of TCP). The results were similar. We also computed the upload bandwidth of the computer by using the *SpeedTest* website (*http://www.speedtest.net*), obtaining a value of approximately *232.500 Bps*. We noticed that the results of our test were closer to *240.000-245.000 Bps*, which was, in fact, the appropriate range for the upload bandwidth of the tested computer. Then, we wanted to decide which packet size is most suitable for estimating the upload bandwidth for the tested computer. We considered that the correct upload bandwidth was *240.000 Bps* and we computed the error between the estimated value and the correct value (i.e. the absolute difference between them) for each of the *6* tests. Then, we multiplied the error by the total generated traffic and we plotted the results in Fig. 3. Some good packet sizes are between *2048* and *16384* bytes; of course, the lower the packet size, the better. We also performed some tests which showed us that the technique is not currently efficient for estimating the available upload bandwidth. We used the same scenario, in which we started two background applications on the tested computer, uploading data at *70 KBps* and, respectively, *80 KBps* overall. The applications were custom made by us. They uploaded random data to a given destination, using $PT \geq 1$ parallel TCP streams each and sending *4 KB* packets. We first set *PT=10* and then we ran the same upload bandwidth estimation tests, except the one with packet size of *1024* bytes. The overall transfer speed of each application decreased by at most *3 KBps* during the tests and the test results were close to *90 KBps* (i.e. the available upload bandwidth was estimated rather accurately). However, when we used *PT=1*, the transfer speeds of the two background applications dropped significantly, depending on the packet size. For packet sizes of *32768* bytes used during the test, the transfer speeds of the applications dropped down to *20-30 KBps* each. Thus, the TCP flows of the background applications can be severely influenced by our proposed technique. If we could somehow instruct the operating system to handle the test packets as low priority packets (i.e. send the test packets only when no other packets are waiting to be sent, or after they were ignored for more than a certain time duration), then we might be able to use only the actual available upload bandwidth. However, it seems that most operating systems consider that every flow has the same priority and packets are sent in a first-come first-served manner. A possible way of estimating the available upload bandwidth *AUB* is the following. We can introduce an upload speed limit $R$ in our technique – thus, peer $S$ will not necessarily upload data continuously. Let $U(R)$ be the upload bandwidth estimation obtained for a limit $R$. If $U(R) \geq cr \cdot R$, then $R \leq AUB$; if $U(R) < cr \cdot R$ then $R > AUB$ (where $0 < cr \leq 1$, but close to *1*). Thus, we could use a search technique (e.g. exponential and binary search) for finding the largest limit $R$ for which $U(R) \geq cr \cdot R$ (i.e. $R \leq AUB$). In the end we mention that our technique also works when the tested machine has multiple physical upload links. In this case we must find a suitable set of helper peers, such that when performing the test, all the upload links are saturated (enough packets are sent through each link), or we could try to test every upload link separately.

We also considered a different approach, for estimating the available upload bandwidth, based on measuring ping times to a set of carefully chosen landmarks from the Internet. The source peer $S$ uploads data at (at most) a (total) given rate $R$ to a subset of helper peers, for a duration $T$, during which it measures the ping times to the set of landmarks. We expect that, as the transfer rate $R$ gets closer to the available upload bandwidth *AUB*, a larger fraction of pings exceed their time limit. Then, we could increase (or decrease) the rate $R$ with small increments, until the ping times satisfy some quality conditions (e.g. a percentage of them are below some threshold), thus converging towards *AUB*. We present below the results of a first set of experiments. Peer $S$ was located in Bucharest, did not have a public IP address, was running Windows Vista and its upload capacity was approx. *60 KBps*. We chose only one helper peer $P$, located at the Politehnica University of Bucharest (UPB), running Linux and having a public IP address. We ran the test scenario *5* times. The maximum transfer rate was limited at: *25 KBps*, *35 KBps*, *40 KBps*, *45 KBps* and *unbounded*. Every time, the total duration of the upload test was *10* minutes. We measured ping times from the peer $S$ to a machine located at the UPB site. Without the test traffic, the ping times ranged from *15* to *60* milliseconds. For the *25 KBps* upper bound, most of the ping times were under *100 msec*, with only *3* occasional ping time spikes (two of which were ping timeouts). For the *35 KBps* limit, most of the ping times were under *400* msec and no ping timeouts occurred. For the *40 KBps*, several pings timed out in the beginning of the test; however, except for this, the ping times were quite constant, not exceeding *500 msec*. For the *45 KBps*, all the ping times during the actual data transfer exceeded our *20* second time limit. In the unbounded case, the average upload rate was *55 KBps* and the ping times showed a steady increase towards our *20* second time out limit, followed by many ping timeouts. From this set of experiments, we draw the following preliminary conclusions. During an upload bandwidth test without variable background traffic, the ping times present quite a regular behavior. We mention that this behavior is also the result of the technique used to limit the transfer rate. We considered several techniques, some of which led to irregular ping time behavior, and we settled on one where the actual upload rate is constantly corrected (both by introducing time delays and by sending at most a number $X$ of bytes at a time, where $X$ depends on the current upload rate and on the total number of bytes transmitted so far). As expected, the average ping time and the median ping time increase with the upper bound of the upload rate. The implemented mechanism is intrusive, because it needs to send a significant amount of extra traffic in the network. However, we believe that it can be used in a useful non-intrusive manner, as follows. In order to estimate *AUB* accurately using this technique, we might need to send data at the same rate as the available bandwidth. We consider this to be too intrusive and we propose the following use in applications *App* which want to use this technique in order to increase their total upload speed. We can estimate if *AUB* is larger than a small value *R* (by sending data at the rate *R* and checking if the ping times satisfy the quality conditions). Let's assume that the current upload rate of *App* is *U*. If *AUB≥R*, we will use the technique again only after the upload transfer rate of *App* becomes *U+R*. Thus, we only generate as much extra traffic as *App* can use. As future work, we intend to find a correlation between a statistical measure *SM* of the ping times and the upload rate *U*. By using the technique for several small values $\{U_1, ..., U_r\}$ of the upload rates and computing the corresponding statistical measures $\{SM_1, ..., SM_r\}$, we hope to find a correlation *U=f(SM)*. Then, by setting an upper limit $SM_{max}$ on the statistical measure, we could compute the largest upload rate *Umax* that we can use.

## 3 Allocating Communication Resources to Customers with Access Restrictions

We consider the following problem. We have *N* communication providers (numbered from *0* to *N-1*), each provider *i* (*0≤i≤N-1*) offering *S(i)* communication resource units. We also have *N* communication resource consumers (also numbered from *0* to *N-1*), each consumer *i* (*0≤i≤N-1*) being able to consume at most *P(i)* resource units. Due to physical (and other) constraints, the resources from a provider *i* (*0≤i≤N-1*) can be allocated only to the consumers *i* and *((i+1) mod N)*. Let's define *ralloc(i,j)* the amount of resources allocated from provider *i* to consumer *j* (*j=i* or *((i+1) mod N)*). These values must be integers and must satisfy the constraints that *ralloc(i,i)+ralloc(i,((i+1) mod N))≤S(i)* and *ralloc(i,i)+ralloc(((i-1+N) mod N), i)≤P(i)* (for every *i*, *0≤i≤N-1*). We want to allocate as many resource units as possible to the consumers, i.e. we want the sum of the *ralloc(\*,\*)* values to be maximum. Actually, we will study a more general function. Let's define the variable *x=ralloc(0,0)* (*0≤x≤XMAX=min{S(0), P(0)}*). Let *rsum(x)* be the maximum sum of the allocated resource units, if *ralloc(0,0)=x*. We want to be able to compute the values of this function for every possible value of *x*. We will start with an *O(N·XMAX)* time algorithm. For every possible value of *x*, we will be able to compute *rsum(x)* in *O(N)* time. We denote this algorithm, returning *rsum(x)*, by *Algo(x)*. We will maintain the values *Palloc(i)*=the number of resource units allocated to consumer *i* and *Salloc(i)*=the number of resource units allocated from the provider *i*. Initially, we will have *Salloc(i)=Palloc(i)=0* (*1≤i≤N-1*) and *Salloc(0)=Palloc(0)=x*. We will traverse the resource providers in order, from *0* to *N-1*. Let's assume that we reached provider *i*. If *i>0* we will try to allocate as many resources as possible from the provider *i* to the consumer *i*. We compute *q=min{S(i)-Salloc(i), P(i)-Palloc(i)}* and allocate *q* resource units from provider *i* to consumer *i*: we set *Salloc(i)=Salloc(i)+q* and *Palloc(i)=Palloc(i)+q*. Afterwards, no matter what the value of *i* is (i.e. for *i=0*, too), we will try to allocate as many resources from provider *i* to the consumer *i'=((i+1) mod N)*. We compute *q'=min{S(i)-Salloc(i), P(i')-Palloc(i')}* and then we allocate *q'* resource units from provider *i* to the consumer *i'*: we set *Salloc(i)=Salloc(i)+q'* and *Palloc(i')=Palloc(i')+q'*. This *O(N)* greedy algorithm allocates the maximum amount of resource units, given that *x* resource units were allocated from provider *0* to the consumer *0*. *rsum(x)* is then equal to the sum of the *Salloc(\*)* (or *Palloc(\*)*) values. The motivation behind this fact is simple. Once *ralloc(0,0)* is fixed, the remaining resources from provider *0* are only useful to the consumer *(0+1) mod N* and, thus, they will be allocated to this consumer. Then, if consumer *(0+1) mod N* can consume any more resources, then it doesn't make sense not to allocate those resources from the provider *(0+1) mod N*, as these resources would be used efficiently. The arguments extend to the other providers and consumers, in increasing order of their index.

In order to improve the time complexity, we will introduce the following functions: *f(i,x)*=the number of resource units allocated from provider *i* to the consumer *i*, given that *x* resources were allocated from provider *0* to consumer *0*, and *g((i+1) mod N, x)*=the number of resources units allocated from provider *i* to the consumer *((i+1) mod N)*, given that *x* resources were allocated from provider *0* to consumer *0*. We will compute these functions one at a time and we will see that they have a very specific structure. We have *f(0,x)=x* (*0≤x≤XMAX*). *g((0+1) mod N, x)=min{S(0)-f(0,x), P(1 mod N)}*. *f(1,x)=min{P(1)-g(1,x), S(1)}*. In general, we have *f(i,x)=min{P(i)-g(i,x), S(i)}* (*1≤i≤N-1*) and *g((i+1) mod N, x)=min{S(i)-f(i,x), P((i+1) mod N}* (*0≤i≤N-2*). *g(0,x)* is defined as *min{S(N-1)-f(N-1,x), P(0)-x}*. These functions can be computed iteratively. We can compute *g(1,\*)* from *f(0,\*)*, then *f(1,\*)* from *g(1,\*)*, then *g(2,\*)* from *f(1,\*)*, then *f(2,\*)* from *g(2,\*)*, and so on (*g(i,\*)* from *f((i-1+N) mod N,\*)* and then *f(i,\*)* from *g(i,\*)*). The important property of these functions is their structure. The values of the *f(i,x)* functions are as follows: for

$0 \leq x \leq A(i,x)$, $f(i,x)=V_1(i,x)$. For $A(i,x) \leq x \leq B(i,x)$, $f(i,x)$ is increasing with slope *1*, i.e. $f(i,x)=V_1(i,x)+(x-A(i,x))$. For $B(i,x) \leq x \leq XMAX$, $f(i,x)=V_2(i,x)$, where $V_2(i,x)=V_1(i,x)+B(i,x)-A(i,x)$. Thus, every function $f(i,x)$ consists of a part where its values are constant, then an increasing part (with slope *1*) and then another part where its values are constant again. Any of these parts can be void. The $g(i,x)$ ($1 \leq i \leq N-1$) functions are similar, except that the values on the middle parts are decreasing, i.e.: their values are constant on an interval $[0,C(i,x)]$, then decreasing (with slope *-1*) on an interval $[C(i,x), D(i,x)]$ and then constant again. $g(0,x)$ is a bit special, in the sense that, at the end, it may contain an extra part where its values are decreasing again (with slope *-1*): thus, its general structure is an interval of constant values, followed by an interval of decreasing values (with slope *-1*), followed by another interval of constant values and, possibly, followed by another interval of decreasing values (also with slope *-1*). Thus, every function *f(i,x)* and *g(i,x)* has a *O(1)* breakpoints. We will sort the coordinates of these breakpoints (including *x=0* and *x=XMAX*) in increasing order and we will generate events for each breakpoint. Each event will have a value: *0*, *+1* or *-1* (depending on whether the corresponding function is constant, increasing or decreasing starting from that breakpoint) and an x-coordinate. We initialize a variable *Sum* as the sum of the values *f(i,0)* and *g(i,0)* ($0 \leq i \leq N-1$), i.e. *rsum(0)=Sum*, and we set $sf(0 \leq i \leq N-1)=sg(0 \leq i \leq N-1)=0$ (*sf(i)* and *sg(j)* will be the current slopes of the functions *f(i,x)* and *g(j,x)*). We also maintain a variable *Dif*, which is initially *0*. Whenever we encounter a new event corresponding to a function *f(i,x)* or *g(i,x)*, we first compute the values of the function *rsum* corresponding to the values *x* between *E'+1* and *E*, where *E'* is the coordinate of the previous event (or *0*). For every value $E'+1 \leq x \leq E$, *rsum(x)* will be equal to *Sum+Dif·(x-E')*. Afterwards, we set *Sum=Sum+Dif·(E-E')* and then we subtract from *Dif* the previous slope of the function (*sf(i)* for *f(i)*, or *sg(i)* for *g(i)*) and add to *Dif* the value associated to the event (then we set *sf(i)* or *sg(i)* to the value associated to the event, depending on the function to which the event corresponds). This way, we can compute all the values of the function *rsum(\*)* in *O(N·log(N)+XMAX)* time. However, we can do even better. A more careful analysis of the structure of the functions *f(\*,x)* and *g(\*,x)* leads to the observation that the function *rsum(x)* has the following structure: it is increasing (with slope *1*) from *x=0* up to *x=U*, then its values are constant up to *x=V* ($V \geq U$) and then its values are decreasing (with slope *-1*) up to *x=XMAX*. Thus, let's assume that $y_1=Algo(0)$ and $y_2=Algo(XMAX)$ are the values computed by the algorithm *Algo* for *x=0* and *x=XMAX*. Based on these values, we will compute $x_1=((y_2-y_1+XMAX)\ div\ 2)$ and $yx_1=Algo(x_1)$. If ($y_2-y_1+XMAX$) is an odd number then we set $x_2=x_1+1$ and $yx_2=Algo(x_2)$; otherwise (if it is even) we set $x_2=x_1$ and $yx_2=yx_1$. Then, we compute $d_1=y_1+x_1-yx_1$ and $d_2=y_2+(XMAX-x_2)-yx_2$. The function *rsum(x)* is increasing (with slope *1*) from *x=0* up to $x=x_1-d_1$ (its values are $rsum(x)=y_1+x$). On the interval $[x=x_1-d_1+1,x=x_2+d_2]$ the function *rsum(x)* is constant (we have $rsum(x)=yx_1$). Finally, *rsum(x)* is decreasing (with slope *-1*) from $x=x_2+d_2+1$ up to *x=XMAX* (its values are $rsum(x)=y_2+(XMAX-x)$). The time complexity of this approach is *O(N+XMAX)*.

## 4 Related Work and Conclusions

Many end-to-end bandwidth estimation tools and techniques have been developed during the past few years, like packet pair/train dispersion, variable packet size, or self-loading periodic streams (see [1] for a short survey on this, and [5]). However, none of them can be used for estimating the upload bandwidth of a machine. Nevertheless, we were inspired by these methods when we developed the upload bandwidth estimation technique presented in this paper. Communication resource allocation problems have been studied in many papers (e.g. [4]), including those presenting video on demand or live streaming applications and models [2, 3].

In this paper we analyzed two problems. One of them is motivated by practical requirements and aims at estimating accurately the upload capacity (total upload bandwidth) of a machine. The other one is interesting from a theoretical point of view and considers the constrained allocation of communication resources to customers. We presented novel algorithmic solutions for both problems. As future work, we will attempt to develop a method for accurately and efficiently estimating the available upload bandwidth of a machine.